\documentclass[journal]{IEEEtran}
\ifx\pdfoutput\undefined
\usepackage{graphicx}
\else
\usepackage[pdftex]{graphicx}
\fi

\begin{document}
\title{Fabrication and characterization
of superconducting circuit QED devices for quantum computation}
\author{Luigi Frunzio, Andreas Wallraff, David Schuster, Johannes Majer, and Robert Schoelkopf% <-this % stops a space
\thanks{Manuscript received October 4, 2004. This work was supported in part by the National Security Agency (NSA)
 and Advanced Research and Development Activity (ARDA) under Army Research Office (ARO) contract number DAAD19-02-1-0045,
  the NSF ITR program under grant number DMR-0325580, the NSF under grant number DMR-0342157, the David and Lucile
  Packard Foundation, the W. M. Keck Foundation.}% <-this % stops a space
\thanks{L. Frunzio is with the Department of Applied Physics, Yale University,
 New Haven, CT 06520, USA (corresponding author phone: 203-432-4268; fax: 203-432-4283; e-mail:
 luigi.frunzio@yale.edu).}
 \thanks{A. Wallraff, D. Schuster, J. Majer, R. Schoelkopf are with the Department of Applied Physics, Yale University,
 New Haven, CT 06520, USA.}}% <-this % stops a space
% note the % following the last \IEEEmembership and also the first \thanks -
% instance, "A\textbf{} \textbf{}B" will typeset as "A B" not "AB". If you want
% "AB" then you have to do: "A\textbf{}\textbf{}B"
\markboth{confirmation no. 1357 Session ID 3EI07}{Shell
\MakeLowercase{\textit{et al.}}: Fabrication and characterization
of superconducting circuit QED devices for quantum computation}
% *** Note that you probably will NOT want to include the author's name in ***
% *** the headers of peer review papers.                                   ***

% If you want to put a publisher's ID mark on the page
% (can leave text blank if you just want to see how the
% text height on the first page will be reduced by IEEE)
%\pubid{0000--0000/00\$00.00~\copyright~2002 IEEE}

%\specialpapernotice{(Invited Paper)}

\maketitle

\begin{abstract}
We present fabrication and characterization procedures of devices
for circuit quantum electrodynamics (cQED). We have made
$3\,\,\mathrm{GHz}$ cavities with quality factors in the range
$10^{4}-10^{6}$, which allow access to the strong coupling regime
of cQED. The cavities are transmission line resonators made by
photolithography. They are coupled to the input and output ports
via gap capacitors. An Al-based Cooper pair box is made by e-beam
lithography and Dolan bridge double-angle evaporation in
superconducting resonators with high quality factor. An important
issue is to characterize the quality factor of the resonators. We
present an RF-characterization of superconducting resonators as a
function of temperature and magnetic field.  We have realized
different versions of the system with different box-cavity
couplings by using different dielectrics and by changing the box
geometry. Moreover, the cQED approach can be used as a diagnostic
tool of qubit internal losses.
\end{abstract}


% Will be ignored for other modes.
\IEEEpeerreviewmaketitle

\section{Introduction}
% \PARstart{A}{demo} file is ....
% \PARstart{A}{}demo file is ....
% \PARstart{T}{his demo} file is ....
\PARstart{W}{e} have recently demonstrated that a superconducting
quantum two-level system can be strongly coupled to a single
microwave photon \cite{Wallraff}. The strong coupling between a
quantum solid state circuit and an individual photon, analogous to
atomic cavity quantum electrodynamics (CQED) \cite{Mabuchi}, has
previously been envisaged by many authors, see \cite{Blais} and
references therein. Our circuit quantum electrodynamics
architecture \cite{Blais}, in which a superconducting charge
qubit, the Cooper pair box (CPB) \cite{Bouchiat}, is coupled
strongly to a coplanar transmission line resonator, has great
prospects both for performing quantum optics experiments
\cite{Walls} in solids and for realizing elements for quantum
information processing \cite{Nielsen} with superconducting
circuits \cite{Nakamura} and also for other architectures
\cite{Iontrap}.

In developing these qubit-resonator systems, one key ingredient is
to design and realize transmission line resonators with high
internal quality factor, $Q_{int}$, and with resonant frequency,
$\nu_{o}$, in the $5-15\,\,\mathrm{GHz}$ range to match the other
energy scales of our device, and to be in the quantum regime
($h\nu_{o} \gg k_{B}T$) at $T=30\,\,\mathrm{mK}$. On the other
hand, the resonator is loaded with input and output capacitances
and we need a loaded quality factor $Q_{L}\approx10^{4}$ in order
to obtain reasonably fast rate of measurement,
$\kappa=\nu_{o}/Q_{L}\approx1\,\,\mathrm{MHz}$ .

In fabricating the transmission line resonator, we opted for a
coplanar waveguide (CPW) for many different reasons. First, a CPW
has a simple layer structure with no need for deposited
insulators. Second, it has a balanced structure with a relatively
easy planar connection to the CPB. Third, a CPW has a $\nu_{o}$
that is relatively insensitive to kinetic inductance and dominated
by geometrical distributed inductance. Last but not the least,
CPW-based structures, made by Al thin film deposited on sapphire,
have been recently shown \cite{Day} to allow very high
$Q\mathrm{'s}$ (order of $10^{6}$).

We decided to fabricate on passivated Si wafers because this is
the substrate on which we had previously developed the qubit
fabrication. We also decided to try as material for the resonators
both Al, for easy compatibility with the qubit process, and Nb,
because its higher critical temperature allows testing of
resonators at higher temperatures.

In section II, we present design consideration for devices for
circuit quantum electrodynamics (cQED). We will show that we can
engineer $Q$ with different coupling of the resonator to the input
and output ports and that the internal losses can be made
negligible at the designed $Q$ \cite{Wallraff}. Section III
introduces the fabrication procedures for both the resonator and
the CPB. Section IV-VI present an RF-characterization of the
superconducting transmission line resonators versus temperature
and magnetic field.

\begin{figure}[h]
\includegraphics[width=3.45in]{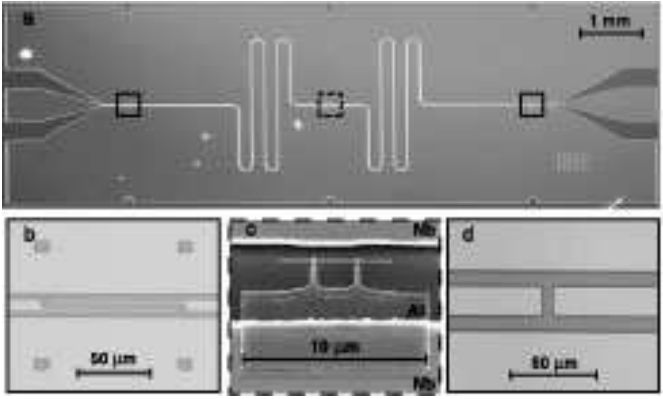}
\label{fig:Sample} \caption{Picture of a device for circuit QED.
(a) The $3\,\,\mathrm{GHz}$ superconducting coplanar waveguide
resonator is fabricated using optical lithography. The length of
the meandering resonator is $l=24\,\,\mathrm{mm}$. The center
conductor is $10\,\,\mathrm{\mu m}$ wide, separated from the
lateral ground planes extending to the edges of the chip by a
$5\,\,\mathrm{\mu m}$ gap. The resonator is coupled by identical
capacitors at each end (solid line squares) to input and output
ports. (b) Micrograph of a coupling capacitance with two
$100\,\,\mathrm{\mu m}$ long and $4\,\,\mathrm{\mu m}$ wide
fingers separated by a $2\,\,\mathrm{\mu m}$ gap. (c) Scanning
electron micrograph of a Cooper pair box fabricated onto the
silicon substrate into the gap between the center conductor (top)
and the ground plane(bottom) in the center of a resonator (dashed
line square) using electron beam lithography and double angle
evaporation of aluminium. (d) Micrograph of a coupling capacitance
with a $4\,\,\mathrm{\mu m}$ gap.}
\end{figure}

\section{Circuit design}
A picture of a $10\times 3\,\,\mathrm{mm^2}$ chip containing a
$3\,\,\mathrm{GHz}$ superconducting Nb CPW resonator is shown in
Fig. 1a. The length of the meandering resonator is
$2l=4\,\,\mathrm{mm}$. The center conductor is $10\,\,\mathrm{\mu
m}$ wide, separated from the lateral ground planes extending to
the edges of the chip by a $5\,\,\mathrm{\mu m}$ gap, resulting in
a wave impedance of the coplanar waveguide of
$Z=50\,\,\mathrm{\Omega}$ to match the impedance of conventional
microwave components. The capacitance per unit length is
$C_{s}\approx 0.13\,\,\mathrm{fF/\mu m^2}$ which gives a total
resonator capacitance of $C=C_{s}l/2=1.6\,\,\mathrm{pF}$. The
resonator is coupled by identical capacitors at each end (see
solid line square in Fig.1a) to an input and output feed line,
fanning out to the edge of the chip and keeping the impedance
constant. In Fig. 1b and 1d are shown micrographs of two of the
coupling capacitors with different geometries. The one in Fig. 1b
consists of two $100\,\,\mathrm{\mu m}$ long and $4\,\,\mathrm{\mu
m}$ wide fingers separated by a $2\,\,\mathrm{\mu m}$ gap. It has
a capacitance, $C_{\kappa ,b}\approx 6\,\,\mathrm{fF}$, larger
than that in Fig.1d, which has a simpler geometry with a
$4\,\,\mathrm{\mu m}$ gap and $C_{\kappa ,d}\approx
0.3\,\,\mathrm{fF}$.

The capacitive coupling to the input and output lines, together
with the loading impedance, $R_{L}=50\,\,\mathrm{\Omega}$, are
very important in determining the loaded quality factor $Q_{L}$,
defined by the following equations:

\begin{equation}\label{1}
    \frac{1}{Q_{L}}=\frac{1}{Q_{int}}+\frac{1}{Q_{ext}}
\end{equation}
where the external quality factor is
\begin{equation}\label{2}
    Q_{ext}=\frac{\omega C}{G_{ext}}
\end{equation}
with
\begin{equation}\label{3}
    G_{ext}=\frac{2R_{L}C_{\kappa}^2\omega ^2}{1+R_{L}^2C_{\kappa}^2\omega ^2}
\end{equation}

There are two possible regimes for the resonator. It can be
undercoupled when $C_{\kappa}$ is small (like $C_{\kappa ,d}$) and
then $Q_{L}\approx Q_{int}$. This is the regime in which it is
possible to measure $Q_{int}$. Otherwise, the resonator can be
overcoupled when $C_{\kappa}$ is large (like $C_{\kappa ,b}$) and
then $Q_{L}\approx Q_{ext}$. It is then possible to engineer the
$Q_{L}$ to obtain fast measurement with $\kappa$ much larger than
the qubit decay rates \cite{Wallraff}.

In Fig. 1c an electron micrograph of a Cooper pair box is shown.
The CPB consists of a $7\,\,\mathrm{\mu m}$ long and
$200\,\,\mathrm{nm}$ wide superconducting island parallel to the
center conductor which is coupled via two $200\times
100\,\,\mathrm{nm^2}$ size Josephson tunnel junctions to a much
larger superconducting reservoir. The CPB is fabricated onto the
silicon substrate (see dashed line square in Fig.1a) in the gap
between the center conductor (top) and the ground plane (bottom)
at an antinode of the electric field in the resonator. The
Josephson junctions are formed at the overlap between the island
and the fingers extending from the reservoir, which is
capacitively coupled to the ground plane. The CPB is a two-state
system described by the hamiltonian
$H=-(E_{el}\sigma_{x}+E_{J}\sigma_{z})/2$ where $E_{el}$ is the
electrostatic energy and $E_{J}=E_{J,max}\cos(\pi\Phi_{b})$ is the
Josephson energy. The overall energy scales of these terms, the
charging energy $E_{el}$ and the Josephson energy $E_{J,max}$, can
be readily engineered during the fabrication by the choice of the
total box capacitance and resistance respectively, and then
further tuned in situ by electrical means. A flux bias
$\Phi_{b}=\Phi/\Phi_{o}$, applied with an external coil to the
loop of the box, controls $E_{J}$. We have demonstrated that
changing the length of the CPB island and its distance to the
center conductor and changing the dielectrics (removing the
passivation step of the Si substrate), we can obtain stronger
couplings of qubit and resonator as predicted by simple
electrostatic calculations of the capacitances.

\section{Device fabrication}
The pattern of 36 different Nb resonators is generated exposing a
bilayer photoresist ($600\,\,\mathrm{nm}$ LOR5A and
$1.2\,\,\mathrm{\mu m}$ S1813) through a mask with traditional UV
photolithography. Then a $200\,\,\mathrm{nm}$ thick Nb film is dc
magnetron sputtered in Ar at $1.5\,\,\mathrm{Pa}$ with a rate of
$1\,\,\mathrm{nm/s}$ in an UHV system with a base pressure of
$20\,\,\mathrm{\mu Pa}$. The substrate is a $2"$
$300\,\,\mathrm{\mu m}$ thick  p-doped (Boron) $(100)$ oriented Si
wafer with resistivity $\rho>1000\,\,\mathrm{\Omega cm}$
previously passivated by thermal wet oxidation with a
$470\,\,\mathrm{nm}$ thick layer of $SiO_{2}$. A lift-off process
in NMP followed by ultrasonic agitation completes the resonator
fabrication.

Al resonators are fabricated on the same type of substrate
depositing a $200\,\,\mathrm{nm}$ thick Al film by thermal
evaporation at a rate of $1\,\,\mathrm{nm/s}$ in the same UHV
system. Then the same mask is used to expose a single photoresist
layer ($1.2$ $\mu m$ S1813) and then realized by wet etching
($8:4:1:1=H_{3}PO_{4}:CH_{3}COOH:HNO_{3}:H_{2}O$) the metal.

In both cases, chips containing individual resonators are obtained
by dicing the Si wafer. The CPB qubit (Fig. 1c) is then fabricated
on an individual resonator by a simple Dolan bridge technique
\cite{Dolan} exposing a bilayer resist ($500\,\,\mathrm{nm}$
MMA-(8.5)MAA EL13 and $100\,\,\mathrm{nm}$ 950K PMMA A3) by e-beam
lithography and then e-beam evaporating Al ($35\,\,\mathrm{nm}$
for the base and $70\,\,\mathrm{nm}$ for the top electrode) at a
rate of $1\,\,\mathrm{nm/s}$ in a double-angle UHV system with a
base pressure of $20\,\,\mathrm{\mu Pa}$. The junction barrier is
realized with a $12\,\,\mathrm{min}$ thermal oxidation in a
$400\,\,\mathrm{Pa}$ of $O_{2}$. A lift-off process in hot acetone
and ultrasonic agitation complete the device. To couple the qubit
reservoir to ground with a large capacitance, the base electrode
is deposited with a little angle taking advantage of the shadow of
the thicker Nb film to define the capacitor.

\begin{figure}[h]
\includegraphics[width=3.45in]{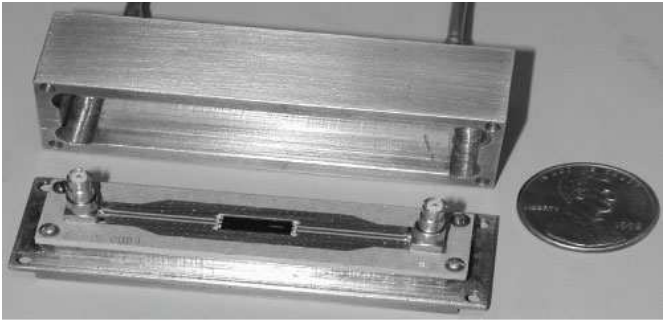}
\label{fig:samplebox} \caption{Picture of the copper sample box
containing a resonator mounted on the PC board.}
\end{figure}

\section{Measurement technique}
The frequency dependence of the transmission through the
resonators \cite{The} was measured using a vector network
analyzer. The equivalent circuit of the measurement setup is shown
in the inset of Fig. 3. The sample was mounted on a PC board in a
closed copper sample box (Fig. 2) equipped with blind mate SMP
connectors that launch the microwaves onto the PC board CPW's. The
sample was cooled to temperatures ranging from the critical
temperature, $T_{c}$ of the superconducting films down to
$T=30\,\,\mathrm{mK}$.

The transmission $S_{21}$ through the resonator around its
fundamental resonant frequency $\nu_{o}$ is shown in Fig. 3 at
$T=30\,\,\mathrm{mK}$.  The curve was acquired using a
$-60\,\,\mathrm{dBm}$ input power \cite{Input} and a room
temperature amplifier. The input power was lowered until no
distortion of the resonance curve due to excessive input power
could be observed. The network analyzer was response calibrated
($S_{21}$) up to the input and output ports of the cryostat and
the absorption of the cabling in the cryostat was determined to be
approximately $-7\,\,\mathrm{dB}$ in a calibrated $S_{11}$ and
$S_{22}$ reflection measurement. The quality factor of the
resonator is determined by fitting a Lorentzian line to the
measured power spectrum as shown by the solid line in Fig. 3. This
is the power spectrum of an undercoupled resonator and from the
fit we have extracted $\nu_{o}= 3.03694\,\,\mathrm{GHz}$. At this
frequency the insertion loss is $L_{o}= -13\,\,\mathrm{dB}$. The
quality factor is determined from the full width at half max of
the fitted power spectrum and is found to be $Q_{L}\approx
Q_{int}=\nu_{o}/2 \delta\nu_{o}=2\pi \nu_{o}/\kappa=0.55\times
10^{6}$.

\begin{figure}[h]
\includegraphics[width=3.45in]{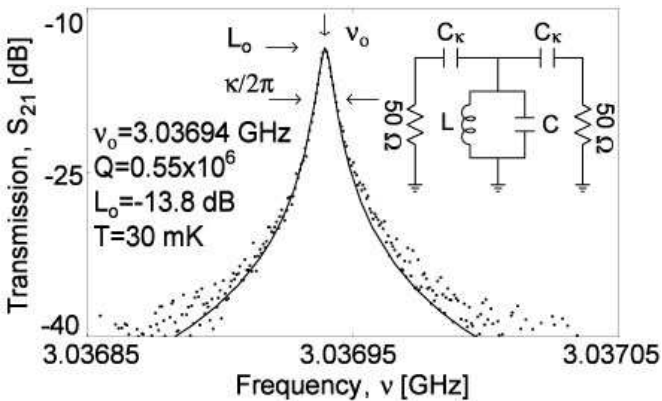}
\label{fig:S21Plot} \caption{Measured transmission power spectrum
of an undercoupled resonator. The solid line is a fit to a
Lorentzian line.}
\end{figure}

\section{Temperature dependence of $Q$ and $\nu_{o}$}
In Fig. 4, we show the measured temperature dependence of the
quality factor $Q$ for an undercoupled resonator (solid dots) and
an overcoupled one (open dots). The lines in Fig. 4 are generated
by summing a $Q_{int}$ that scales exponentially with the reduced
temperature, $T_{c}/T$, in parallel with a constant $Q_{ext}$. At
low temperature, the coupling saturates the $Q$ of the overcoupled
resonator, while it seems that $Q$ for the undercoupled one has
still some weak temperature dependence whose nature is still
unknown. We speculate that either vortices or losses in the
dielectrics could limit the $Q$ of this resonator but neither of
these interpretations offer an easy understanding of the weak
temperature dependence.

\begin{figure}[h]
\includegraphics[width=3.45in]{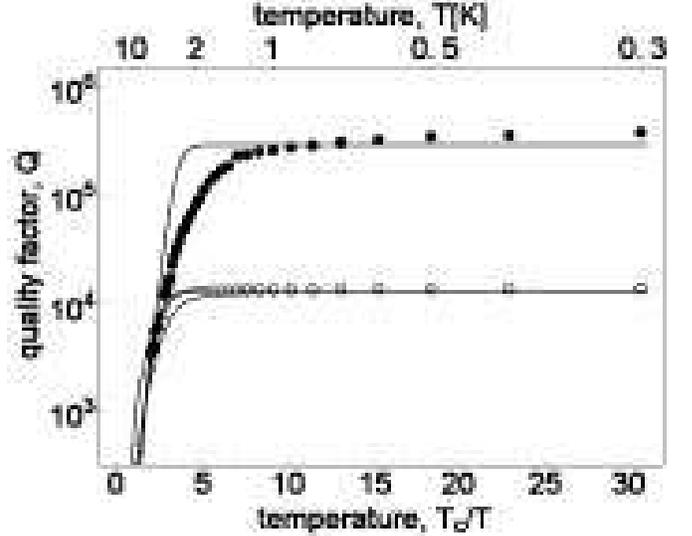}
\label{fig:nuvsT} \caption{Temperature dependence of the quality
factor $Q$ of two $3\,\,\mathrm{GHz}$ superconducting Nb coplanar
waveguide resonators at their first harmonic resosonant frequency
($6\,\,\mathrm{GHz}$). Solid dots are data collected on a
undercoupled resonator and open dots are from an overcoupled one.
The lines are generated by summing a $Q_{int}$ that scales
exponentially with the reduced temperature, $T_{c}/T$, in parallel
with a constant $Q_{ext}$}
\end{figure}

\begin{figure}[h]
\includegraphics[width=3.45in]{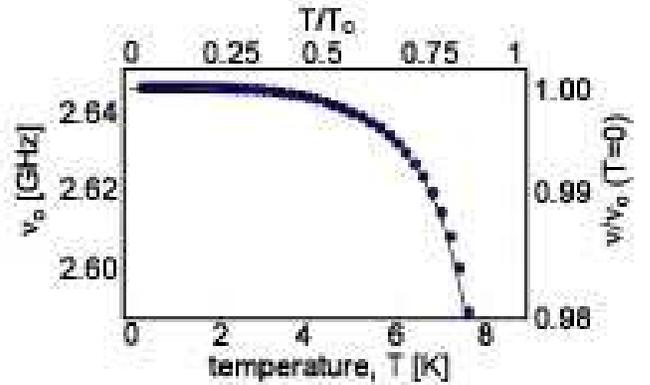}
\label{fig:nuvsT} \caption{Temperature dependence of the resonant
frequency $\nu_{o}$ of a superconducting Nb coplanar waveguide
resonator. Solid line is a fit to a kinetic inductance model.}
\end{figure}

We have observed a shift of the resonant frequency $\nu_{o}$ with
temperature as shown in Fig. 5, which can be understood in terms
of the temperature dependent kinetic inductance of the resonator
\cite{Day, Yoshida}. $\nu_{o}$ is proportional to $1/\sqrt{L}$,
where the total inductance of the resonator $L$ is the sum of the
temperature independent geometric inductance $L_{m}$ and the
temperature dependent kinetic inductance $L_{k}$. The kinetic
inductance scales as $L_{k}\propto \lambda_{L} (T)^2$, where
$\lambda_{L} (T)$ is the temperature dependent London penetration
depth. The best fit in Fig. 4 was achieved for a ratio
$L_{k}/L_{m}\approx 4\%$ and a critical temperature of
$T_{c}\approx 8.75$ $K$, which we have independently measured on a
test sample fabricated on the same wafer.

\begin{figure}[h]
\includegraphics[width=3.3in]{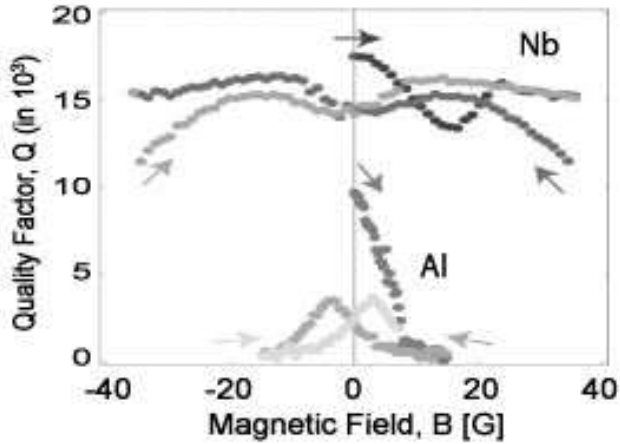}
\label{fig:QvsB} \caption{Magnetic field dependence of the quality
factor $Q$ of two different superconducting coplanar waveguide
resonators at $T=300\,\,\mathrm{mK}$. In the upper part data refer
to a Nb resonator, while in the lower part they refer to an Al
resonator. Arrows indicate the direction in which the magnetic
field was swept in both case starting from zero.}
\end{figure}

\section{Magnetic field dependence of $Q$}
As explained in section II, we need to apply a magnetic field
perpendicular to the qubit loop in order to tune $E_{J}$. Then, we
measured the quality factor of two resonators as a function of the
magnetic field at $T=300\,\,\mathrm{mK}$, as shown in Fig. 6. It
is evident that the Nb film (upper part) is less sensitive to the
applied field than the Al film (lower part). In both cases there
seems to be a reproducible and irreversible hysteretic behavior
that can be reset by thermal cycling the sample. In our recent
works \cite{Wallraff}, we have observed a focusing effect on the
magnetic field such that the effective field in the gap of the
resonator was approximately two orders of magnitude larger than
the applied magnetic field. We believe that the hysteretic
phenomena could be in fact a result of vortices being trapped in
the resonator film due to these large effective fields.

\section{Conclusion}
In summary, we have designed and fabricated devices for realizing
a circuit quantum electrodynamics architecture in which a qubit
can be strongly coupled to a single microwave photon. We have
shown that we can engineer $Q$ with different coupling of the
resonator to the input and output ports and that the internal
losses can be made negligible at the designed $Q$. Indeed, we have
achieved high $Q=0.55\times 10^{6}$ in the undercoupled CPW
resonators and $Q\approx 10^{4}$ in the overcoupled ones, which
allow fast measurement of the qubit.

To help determine the mechanism of the losses, one can fabricate
resonators on different substrates (Si with different resistivity,
sapphire, $\mathrm{Si}_{3}\mathrm{N}_{4}$)), or in different
superconductors (Ta, Al). In fact, quality factor measurements in
this type of resonant circuits serve as a sensitive probe of
material losses in dielectrics and superconductors in the GHz
frequency range at millikelvin temperatures. These presently
unknown properties may in fact pose a serious limit for all
superconducting qubits, though the large internal $Q\mathrm{'s}$
already observed are highly encouraging. Better knowledge of the
material losses, and techniques to characterize them, may be
crucial not only for future improvements of circuit QED devices,
but also for designing and realizing robust, long-lived
superconducting qubits.


\begin{thebibliography}{9}

\bibitem{Wallraff}
A. Wallraff, D. Schuster, A. Blais, L. Frunzio, R.-S. Huang, J.
Majer, S. Kumar, S. Girvin, and R. Schoelkopf, "Strong coupling of
a single photon to a superconducting qubit using circuit quantum
electrodynamics," Nature 431, 162-167 (2004), (also arXiv:
cond-mat/0407325).

D. Schuster, A. Wallraff, A. Blais, L. Frunzio, R.-S. Huang, J.
Majer, S. M. Girvin, and R. J. Schoelkopf, "AC-Stark shift and
dephasing of a superconducting quibit strongly coupled to a cavity
field," Phys. Rev. Lett., to be published (2004)(also arXiv:
cond-mat/0408367).

\bibitem{Mabuchi}
H. Mabuchi and A. Doherty, "Cavity quantum electrodynamics:
Coherence in context," Science 298, 1372-1377 (2002).

\bibitem{Blais}
A. Blais, R.-S. Huang, A. Wallraff, S. Girvin, and R. Schoelkopf,
"Cavity quantum electrodynamics for superconducting electrical
circuits: an architecture for quantum computation," Phys. Rev. A
69, 062320 (2004).

\bibitem{Bouchiat}
V. Bouchiat, D. Vion, P. Joyez, D. Esteve, and M. H. Devoret,
"Quantum coherence with a single Cooper pair," Physica Scripta
T76, 165-170 (1998).

\bibitem{Walls}
D. Walls and G. Milburn, \emph{Quantum optics}, Berlin,
Spinger-Verlag, 1994.

\bibitem{Nielsen}
M. A. Nielsen and I. L. Chuang, \emph{Quantum computation and
quantum information}, Cambridge University Press, 2000.

\bibitem{Nakamura}
Y. Nakamura, Y. A. Pashkin, and J. S. Tsai, "S. Coherent control
of macroscopic quantum states in a single- Cooper-pair box,"
Nature 398, 786-788 (1999). D. Vion, A. Aassime, A. Cottet, P.
Joyez, H. Pothier, C. Urbina, D. Esteve, and M.H. Devoret,
"Manipulating the quantum state of an electrical circuit," Science
296, 886-889 (2002). J. M. Martinis, S. Nam, J. Aumentado, and C.
Urbina, "Rabi oscillations in a large Josephson-junction qubit,"
Phys. Rev. Lett. 89, 117901 (2002). Y. Yu, S. Han, X. Chu, S.-I.
Chu, Y. Wang, "Coherent temporal oscillations of macroscopic
quantum states in a Josephson junction," Science 296, 889-892
(2002). I. Chiorescu, Y. Nakmura, C. J. P. M. Harmans, and J. E.
Mooij, "Coherent quantum dynamics of a superconducting flux
qubit," Science 299, 1869-1871 (2003). T. Yamamoto, Y.A. Pashkin,
O. Astafiev, Y. Nakamura, J.S. Tsai, "Demonstration of conditional
gate operation using superconducting charge qubits," Nature 425,
941-944 (2003). I. Chiorescu, P. Bertet, K. Semba, Y. Nakmura, C.
J. P. M. Harmans, and J. E. Mooij, "Coherent dynamics of a flux
qubit coupled to a harmonic oscillator," Nature 431, 159-162
(2004).

\bibitem{Iontrap}
A. S. Sørensen, C. H., van derWal, L. Childress, and M. D. Lukin,
"Capacitive coupling of atomic systems to mesoscopic conductors,"
Phys. Rev. Lett. 92, 063601 (2004). L. Tian, P. Rabl, R. Blatt,
and P. Zoller, "Interfacing quantum optical and solid state
qubits," unpublished (also arXiv: quant-ph/0310057).

\bibitem{Day}
P. K. Day, H. G. LeDuc, B. A. Mazin, A. Vayonakis, and J.
Zmuidzinas, "A broadband superconducting detector suitable for use
in large arrays," Nature 425, 817–821 (2003).

\bibitem{Dolan}
G. J. Dolan, "Offset masks for lift-off processing," Appl. Phys,
Lett. 31, 337-339 (1977).

\bibitem{The}
The transmission is measured in $dB=10\log |V_{2}/V_{1}|^{2}$,
where $V_{2}$ is the voltage measured at the input port of the
analyzer and $V_{1}$ is the voltage applied at the output port of
the analyzer.

\bibitem{Input}
The input power is in $dBm$ where $-60$ $dBm=20\log (1\mu W/1
mW)$.

\bibitem{Yoshida}
K. Yoshida, K. Watanabe, T. Kisu, and K. Enpuku, "Evaluation of
magnetic penetration depth and surface resistance of
superconducting thin films using coplanar waveguides," IEEE Trans.
on Appl. Supercond. 5, 1979-1982 (1995).

\end{thebibliography}
\end{document}